\documentclass[journal=jacsat,manuscript=article]{achemso}

\usepackage[version=3]{mhchem} 
\usepackage{subfigure}

\usepackage{epstopdf}

\author{Markus Rollinger}
\affiliation{Fachbereich Physik and Forschungszentrum OPTIMAS, Technische Universit\"{a}t Kaiserslautern, 67663 Kaiserslautern, Germany}
\author{Philip Thielen}
\affiliation{Fachbereich Physik and Forschungszentrum OPTIMAS, Technische Universit\"{a}t Kaiserslautern, 67663 Kaiserslautern, Germany}
\alsoaffiliation{Graduate School of Excellence Materials Science in Mainz, Gottlieb-Daimler-Str. 47, 67663 Kaiserslautern, Germany}
\author{Emil Melander}
\affiliation{Department of Physics and Astronomy, Uppsala University, Box 516, SE-75120, Uppsala, Sweden}
\author{Erik \"{O}stman}
\affiliation{Department of Physics and Astronomy, Uppsala University, Box 516, SE-75120, Uppsala, Sweden}
\author{Vassilios Kapaklis}
\affiliation{Department of Physics and Astronomy, Uppsala University, Box 516, SE-75120, Uppsala, Sweden}
\author{Bj\"{o}rn Obry}
\affiliation{Fachbereich Physik and Forschungszentrum OPTIMAS, Technische Universit\"{a}t Kaiserslautern, 67663 Kaiserslautern, Germany}
\author{Mirko Cinchetti}
\affiliation {Fachbereich Physik and Forschungszentrum OPTIMAS, Technische Universit\"{a}t Kaiserslautern, 67663 Kaiserslautern, Germany}
\author{Antonio Garc\'{\i}a-Mart\'{\i}n}
\affiliation{IMM-Instituto de Microelectronica de Madrid (CNM-CSIC), Isaac Newton 8, PTM, Tres Cantos, E-28760 Madrid, Spain}
\author{Martin Aeschlimann}
\affiliation {Fachbereich Physik and Forschungszentrum OPTIMAS, Technische Universit\"{a}t Kaiserslautern, 67663 Kaiserslautern, Germany}
\author{Evangelos Th. Papaioannou}
\email{papaio@rhrk.uni-kl.de}
\affiliation{Fachbereich Physik and Forschungszentrum OPTIMAS, Technische Universit\"{a}t Kaiserslautern, 67663 Kaiserslautern, Germany}

\title
  {Light localization and magneto-optic enhancement in Ni anti-dot arrays} 

\abbreviations{IR,NMR,UV}

\begin{document}

\begin{abstract}
The excitation of surface plasmons in magnetic nano-structures can strongly influence their magneto-optical properties. Here, we use photoemission electron microscopy to map the spatial distribution of the electric near-field on a nano-patterned magnetic surface that supports plasmon polaritons. By using different photon energies and polarization states of the incident light we reveal that the electric near-field is either concentrated in spots forming a hexagonal lattice with the same symmetry as the Ni nano-pattern or in stripes oriented along the $\Gamma$-K direction of the lattice and perpendicular to the polarization direction. We show that the polarization-dependent near-field enhancement on the patterned surface is directly correlated to both the excitation of surface plasmon polaritons on the patterned surface as well as the enhancement of the polar magneto-optical Kerr effect. 

\end{abstract}

\section{Introduction}
The field of magneto-plasmonics is growing rapidly nowadays with the main aim to explore the combination of magnetic and plasmonic functionalities in patterned nano-structures~\cite{Antonio}.  The presence of magnetic materials in plasmonic structures offers the possibility to influence plasmonic resonances with the magnetization and an external magnetic field.  The idea of controlling  plasmonic resonances with magnetism has already been proposed for  applications in the field of bio-sensing~\cite{bio-sensing15}. Furthermore, the magnetic field can be used to modify the dispersion relation of surface plasmon polaritons (SPPs) as it has recently been shown in an interferometer-like structure composed of trilayers of Au/Co/Au~\cite{Temnov2010} with possible applications in the area of telecommunications. 

\noindent On the other hand, the presence of surface plasmon polaritons in magnetic materials can also strongly modify the magneto-optical response. In particular, the excitation of surface plasmon polaritons has been found to enhance the polar magneto-optical Kerr effect (P-MOKE) signal in Ni~\cite{melander:063107, Papaioannou:11,Hui15},  Fe~\cite{PhysRevB.81.054424}, and Co~\cite{Ctistis2009,smirnov2010,Lee2012} anti-dot films.  Moreover, significant enhancement has been demonstrated in hybrid structures composed of  noble and magnetic metals/dielectrics such as Au/Co and Au/Iron garnets~\cite{martin-becerra:183114, Papaioannou15, Belotelov, Caballero:15} and multilayers like Co/Pt~\cite{Xia2009}. SPPs have also been  shown to influence the transverse magneto-optic response (T-MOKE)~\cite{melander:063107,papa2010,adeyeye}.

\noindent Recently it has been reported that Ni nano-islands act as a magneto-plasmonic material, being able to sustain localized surface plasmons (LSPs)~\cite{Bonanni2011}. It was shown that the polarizability of LSPs can invert the sign of the magneto-optical Kerr rotation. Furthermore, Rubio-Roy et al.~\cite{Rubio2012} studied SPPs and LSPs in a Au nano-hole array filled with iron oxide nanoparticles and Ni nanoparticles. They measured an enhanced magneto-optical activity up to an order of magnitude for wavelengths that are correlated to the excitation of localized surface plasmons. They attributed the observed magneto-optical response to the increase of the polarization conversion efficiency, and they found the contribution of reflectance modulations to be negligible. Macaferri et al.~\cite{Maccaferi13} showed that a localized surface resonance excitation transverse to the electric field of the incident light induces the observed shape of the magneto-optic spectrum. Furthermore, Kataja et. al.~\cite{Kataja15} extended the previous studies on isolated and randomly arranged magnetic particles to periodically arranged magnetic Ni nanoparticles in a two-dimensional rectangular lattice. They have shown that the nanoparticle arrays exhibit strong magneto-optical response in the presence of LSPs that is determined by the lattice period parallel to the electric driving field opposite to the purely optical response which is dominated by the period orthogonal to the driving field. The magneto-plasmonic structures have  in addition initiated studies in the nonlinear optical regime. Excitation of SPPs at an interface of a Co anti-dot film was shown to not only enhance the second-harmonic generation (SHG) efficiency, but also to activate a quadrupole nonlinear-optical mechanism~\cite{PhysRevB.88.075436} while the control of magnetic contrast in single crystalline Fe films was achieved with nonlinear magneto-plasmonics~\cite{Lupke14}.

\noindent In spite of these numerous studies on the plasmonic behavior of magneto-plasmonic lattices and the enhanced magneto-optical signal, the mechanism of coupling between plasmonic and magneto-optical properties is not fully clarified yet. Here, we tackle this open issue by combining photoemission electron microscopy (PEEM) with magneto-optical spectroscopy.
Using PEEM, we can visualize the distribution of the electric near-field on the patterned surface with a spatial resolution below the optical diffraction limit. We find remarkably different polarization dependent PEEM images that are recorded at photon energies where SPPs are expected. Comparing the PEEM images to the magneto-optical spectra we unveil a direct correlation between the near-field distribution of the SPPs and the enhancement of the magneto-optical response.

\section{Results and discussion}

\subsection{Reflectivity and plasmonic resonances}

\noindent In the plasmonic research field, the symmetry of the unit cell of a patterned structure plays a central role  since it defines the position of the plasmonic resonances. In the case of  metallic anti-dot arrays, which are studied here, their capability to sustain  SPPs that are responsible for the  enhanced optical transmission is mainly defined by the geometry of the unit cell~\cite{Ebbesen1998}. SPPs have higher in-plane momentum than the incident light. However, it is possible to excite SPPs by photons of a specific energy using a grating structure (such as an anti-dot lattice), that provides the needed additional momentum to compensate for the momentum mismatch~\cite{Maier2007,ebbesen07}. 

\begin{figure}[]
 \begin{center}
 \includegraphics[width=10 cm]{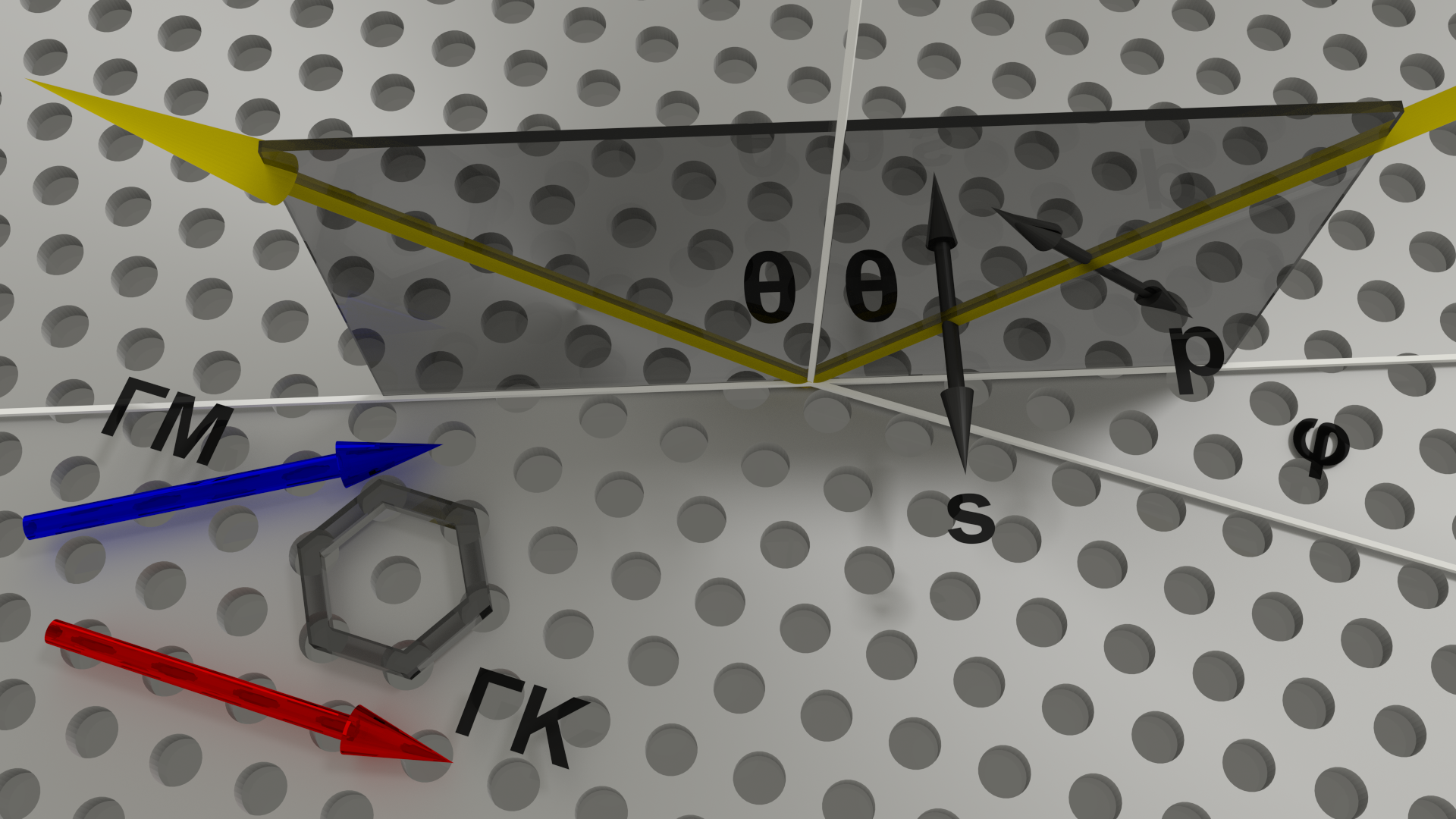}
  \end{center}
  \caption{Schematic of the used geometry and the definition of the angles. $\theta$ is the angle of incidence, $\phi$ the angle between the polarization plane and the nearest neighbor hole ($\Gamma$-K), $\Gamma$-K is the nearest  and $\Gamma$-M next-nearest neighbor direction noted in the real space. All of our measurements were performed by keeping $\phi=0$. S- and p-polarized light correspond to polarization directions perpendicular and parallel to the plane of incidence respectively. Having $\phi=0$ and small $\theta$, s-polarization is equivalent to the alignment of polarization vector along the $\Gamma$-M direction and p-polarized light corresponds to polarization direction along the $\Gamma$-K direction.}
\label{fig:antidot}
\end{figure}

\noindent In our study, we investigate a Ni anti-dot film of 100 nm thickness that was grown on a Si substrate on top of a Ti seed layer of $ 2 \ \textrm{nm}$ thickness. Self-assembly nano-sphere lithography with polystyrene spheres leads to a hexagonal hole pattern\cite{Papaioannou:11} with holes of  $d = 275 \ \textrm{nm}$ diameter and a center to center distance of $a = 470\ \textrm{nm}$, see Fig.~\ref{fig:antidot}. The Ni film is covered by a $ 2 \ \textrm{nm}$ thick protective Au layer. The schematic representation of the hexagonal antidot pattern of Fig.~\ref{fig:antidot}  defines all directions and angles used in this work. $\theta$ is the angle of incidence, the arrow $\Gamma$-K is the nearest  and the arrow $\Gamma$-M the next-nearest neighbor hole direction noted on the schematic for the real space, $\phi$ is the angle between the polarization plane and the nearest neighbor ($\Gamma$-K) direction (in all of our experiments we kept $\phi=0$), and  s- and p-polarized light correspond to polarization states perpendicular and parallel to the plane of incidence respectively. Due to  $\phi=0$ and the very small $\theta$, s-polarization is equivalent to the alignment of polarization vector along the $\Gamma$-M direction, and p-polarization is equivalent to alignment along  the $\Gamma$-K direction.

\noindent Reflectivity spectra were recorded  with different polarization states of light for the Ni anti-dot sample and compared to  a reference continuous film of exactly the same composition and thickness. The reflectivity measurements were recorded at angle of incidence $\theta=6.5^\circ$  and without application of a magnetic field. While keeping $\phi=0^\circ$, the polarization direction was initially aligned horizontally along the $\Gamma$-K direction  (Fig.~\ref{fig:antidot}). The light polarization was then rotated  $90^\circ$, which is aligned along the $\Gamma$-M direction.

\begin{figure}[]
 \begin{center}
 \includegraphics[width=8 cm]{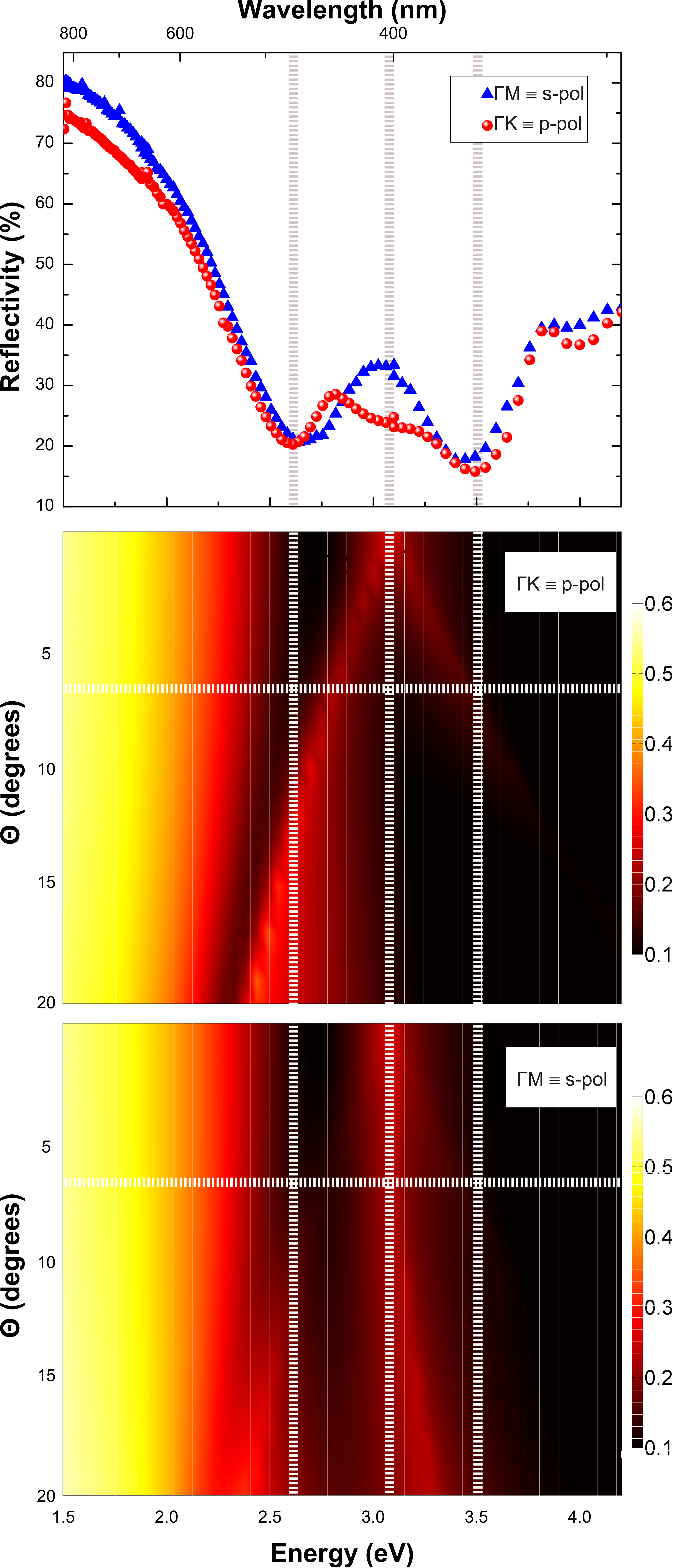}
  \end{center}
  \caption{(Upper panel) Experimental reflectivity curves as a function of energy for the anti-dot Ni based film. Measurements were performed with incident light polarized along the nearest neighbor ($\Gamma$-K) and next-nearest neighbor ($\Gamma$-M) direction. The data have been normalized to the reflectivity of a reference continuous film of the same composition. (Lower panels) Calculated  reflectivity at the Ni-air interface as a function of photon energy and angle of incidence for polarization along $\Gamma$-K and $\Gamma$-M. The plane of incidence is set at $\phi = 0$ parallel to the $\Gamma$-K direction of the hexagonal lattice. Horizontal lines indicate the experimental angle of incidence $\theta$ and the perpendicular lines the energy position of the resonances.}
\label{fig:reflectivity}
\end{figure}

\noindent Figure \ref{fig:reflectivity} (top graph) reveals relative reflectivity curves for incident light polarized along the $\Gamma$-K and $\Gamma$-M direction for the patterned sample. The reflectivity is normalized relative to the reflectivity of the continuous reference film. Drops in the reflectivity of the patterned sample are observed for light polarized along both the $\Gamma$-K and $\Gamma$-M direction at specific energies. The minima of reflectivity are correlated to surface plasmon polariton excitations at the metal/air interface. The condition for resonant excitation of SPPs is given by the momentum matching between SPP, incident light and lattice geometry:

\begin{equation}
\label{eq:matching}
 \overrightarrow{\bf k}_{SPP} = \overrightarrow{\bf k}_{x} \pm  i \overrightarrow {\bf G}_{x} \pm j \overrightarrow {\bf G}_{y}
\end{equation}

\noindent where $\overrightarrow {\bf {\textrm{k}}}_{SPP}$ is the surface plasmon polariton wave vector, $\textrm{k}_{x} =  \textrm{k}_{0} \sin\theta$ is the component of the incident wave vector that lies in the plane of the sample, $ \overrightarrow {\bf G}_{x}$, and $\overrightarrow {\bf G}_{y}$ are the basis vectors of the reciprocal hexagonal lattice and i, j are integers.  Such coupling gives rise to so-called Bragg surface plasmons~\cite{PhysRevB.85.245103, PhysRevB.74.245415, nl0710506}. Figure ~\ref{fig:reflectivity} (lower graphs) shows numerical simulations of the reflectivity obtained for light polarized along the $\Gamma$-K and $\Gamma$-M direction for a Ni anti-dot film calculated by using the scattering matrix approach adapted to deal with arbitrary orientations of magnetization~\cite{PhysRevB.85.245103}. SPP resonances correspond to regions with low reflectivity (dark color) and appear at energies close to the onset of diffraction (bright color). The drawn lines indicate the intersection of measured reflectivity minima (lines parallel to $\Theta$-axis) with the calculated reflectivity for the given angle of incidence of our experiment (lines parallel to Energy-axis). The  good agreement between theory and experiment confirms that the reflectivity minima are due to the excitation of surface plasmons on our magneto-plasmonic structure. 

\subsection{Photoemission electron microscopy on the magneto-plasmonic structure}

\noindent To provide direct evidence for the excitation of SPPs on magneto-plasmonic structures we use Photoemission Electron Microscopy (PEEM). PEEM is a powerful technique for imaging the photoelectron distribution of the sample on a nanometer local scale \cite{Cinchetti2005,Vogelgesang2010,Aeschlimann2010}. More precisely, the electrostatic lens system offers a spatial resolution of 25 to 30~nm, beyond the optical diffraction limit, and is well-suited to image the distribution of the electric near-field within our nm-scaled anti-dot structure. 
As light sources for the microscope we use a mode-locked Ti:Sapphire laser oscillator with a photon energy of $ \hbar \omega = 1.55$~eV and its frequency-doubled fundamental mode with a photon energy of $ \hbar \omega = 3.10$~eV as well as a commercial optical parametric oscillator that provides variable photon energies between 1.55~eV and 3.59~eV. It is worth noting here that the angle of incidence of the laser light impinging on the sample surface was $\theta=4^\circ$ relative to the surface normal, that is equivalent to the one used for the polar Kerr rotation measurements. This will allow us later to directly compare the PEEM experiment with the MOKE signal of the sample.

\begin{figure}[]
 \begin{center}
 \includegraphics[width=16 cm]{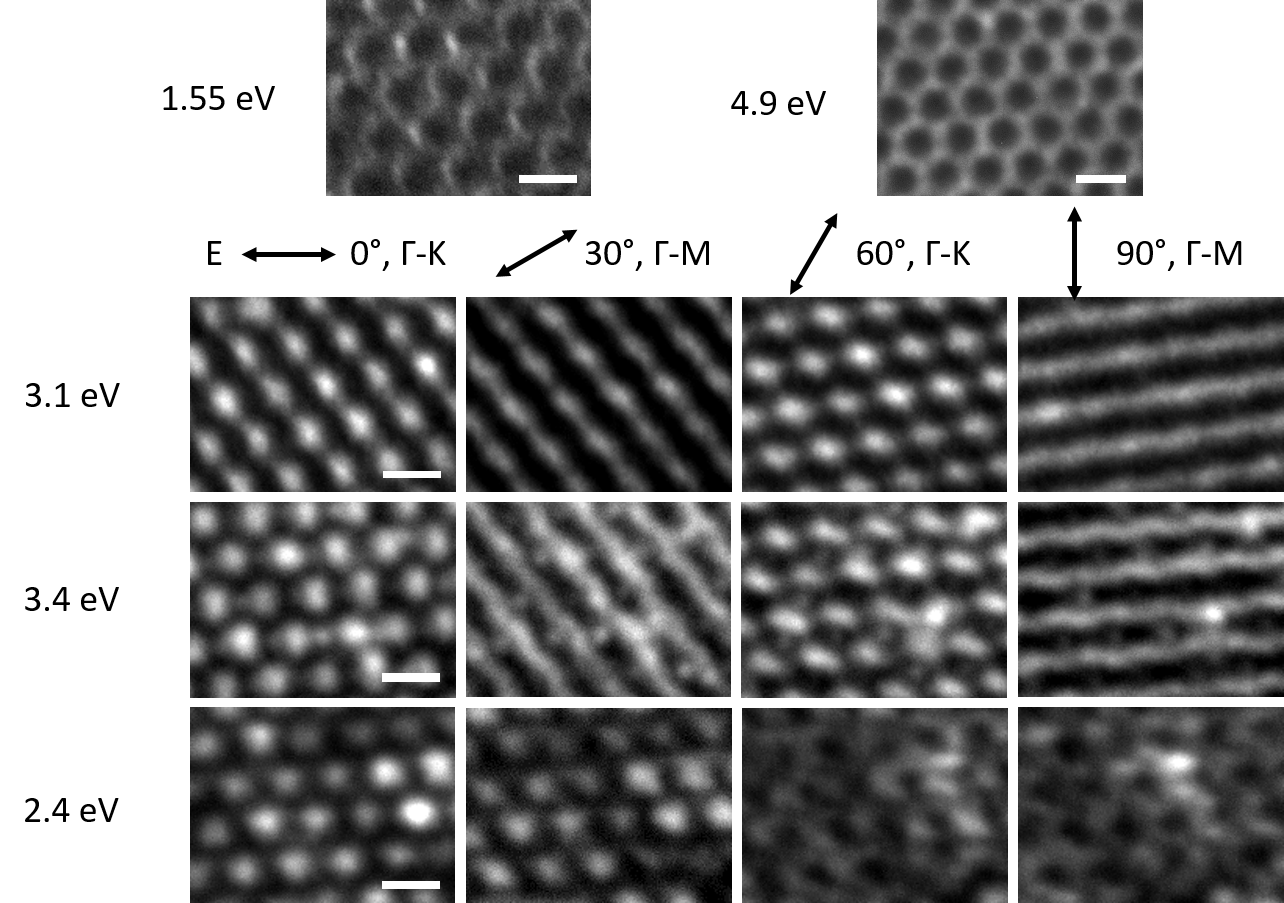}
  \end{center}
  \caption{ PEEM images recorded at different photon energies of the Ni based anti-dot film. The scale bar for all images is 3 $\mu$m. (First row) Images of the patterned structure by using  the excitation with linearly polarized laser light of $ \hbar \omega = 1.55$~eV, as well as with a mercury discharge lamp at $ \hbar \omega =4.9$~eV. (Second row) Photoemission patterns excited at $ \hbar \omega = 3.1$~eV. Spot-like emission is observed from the region in between the holes for incident light aligned along the $\Gamma$-K direction. Incident light polarized along the $\Gamma$-M direction causes emission patterns that form lines perpendicular to the E-field vector direction. (Third row)  Photoemission pattern excited at $ \hbar \omega = 3.4$~eV are similar to the patterns observed for $\hbar \omega = 3.1$~eV. (Fourth row) Photoemission pattern excited at $ \hbar \omega = 2.4$~eV. Weak spot-like emission as well as mainly homogeneous emission is observed that does not follow the symmetry of the patterned while rotating the polarization.}
\label{fig:peem}
\end{figure}

\noindent In Fig.~\ref{fig:peem} we present PEEM images recorded at different excitation energies. The excitation of SPPs due to the coupling of the incident light with the periodic lattice will cause an enhancement of the electric near-field that can be directly measured with PEEM. The top two images of Fig.~\ref{fig:peem} are taken at a photon energy of $\hbar\omega=1.55$~eV with the Ti:Sa oscillator and with $\hbar\omega=4.9$~eV from a mercury discharge lamp. Since at these photon energies no SPP excitation is expected, we take these images as reference, since they only reveal the topography of the sample. The next three rows of Fig.~\ref{fig:peem} show images of the detected electron intensity for different polarization directions, i.e. aligned along the $\Gamma$-K direction ($0^\circ$ and $60^\circ$) and along the $\Gamma$-M direction ($30^\circ$ and $90^\circ$). They are recorded at photon energies of $\hbar\omega=3.1$~eV and $3.4$~eV, i.e.~close to the onset of SPP resonances and at a photon energy of $\hbar\omega=2.4$~eV.

\noindent At $ \hbar \omega = 3.1$~eV PEEM images show very specific emission patterns for different polarizations: either spot-like or continuous lines. In particular, when the incident E-field is aligned along the direction of the nearest neighbors, PEEM images (for $\Gamma$-K direction) reveal dot-like emission from the sample surface. The spots are formed between the holes and they are also arranged in a hexagonal pattern. This behavior is repeated for every 60$^\circ$ of rotation of the polarization state of light, thus every time the polarization is oriented along the $\Gamma$-K direction of the hexagonal hole lattice. When the E-field is rotated by $30^\circ$ or $90^\circ$, $\Gamma$-M direction, strong changes in the PEEM images are observed. Now bright continuous lines are formed perpendicular to the polarization direction of light. Again the lines appear on the Ni and not in the holes. The pattern is also repeated for every 60$^\circ$ of rotation of the polarization of the light, mirroring the symmetry of the structure for the $\Gamma$-M direction. At $\hbar \omega = 3.4$~eV photon energy the PEEM images exhibit a similar behavior: hexagonal bright spot formation for incident light polarized along the $\Gamma$-K direction and continuous lines for incident light polarization along the $\Gamma$-M direction.

\noindent The results for a photon energy of $\hbar \omega = 2.4$~eV are obtained at the outer edge of a SPP resonance. Spot-like emission is observed from different areas of the surface for $0^\circ$ and $30^\circ$ rotation of the polarization of the incident light while a more homogeneous emission is seen for $60^\circ$ and $90^\circ$ of rotation. The small changes in the photoemission patterns are not as distinct as before and evenmore do not mirror the rotational symmetry of the lattice of 60$^\circ$. A distinct line formation in the photoemission pattern is not discernible here for any angle of polarization.

\begin{figure}[]
 \begin{center}
 \includegraphics[width=16 cm]{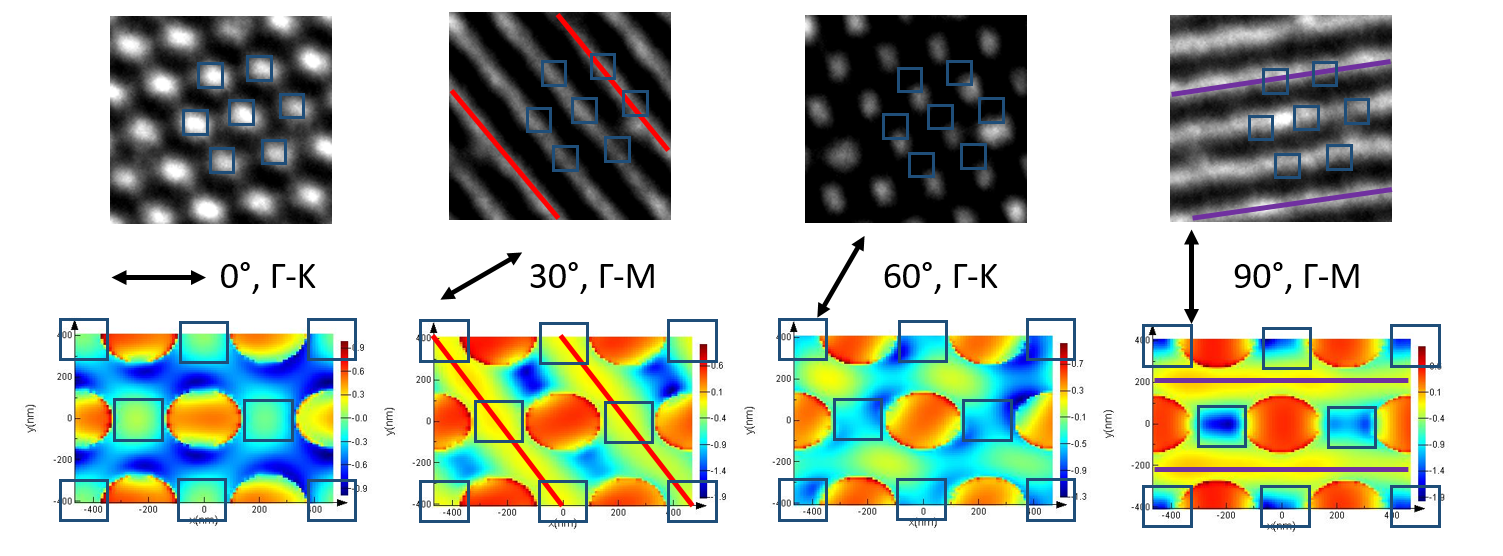}
  \end{center}
  \caption{(Upper panel) Experimental photoemission patterns excited at $ \hbar \omega = 3.1$~eV  for different polarization states as indicated. (Lower panel) Simulated near-field distribution for different polarization states. A very good agreement between the PEEM measurements and the near-field simulations is observed.}
\label{fig:simulation}
\end{figure}

\noindent In order to understand the origin of the photoelectrons and thus the near-field distribution of the exciting light in the PEEM measurements, we compare the measured photoemission pattern to near-field simulations. Those simulations are performed with a commercial-grade simulator based on the finite-difference time-domain method (Lumerical FDTD Solutions, www.lumerical.com). The design of the simulated structure is the same as the magneto-plasmonic sample accounting for the layered structure including substrate, seed layer, Ni film and Au capping layer. As excitation source for the simulations shown here we use a plane-wave source with $ \hbar \omega = 3.1$~eV photon energy. The field distribution is recorded within the 2~nm thick Au capping layer.
Figure \ref{fig:simulation} compares the experimental photoemission patterns for different polarization angles measured at a photon energy of $\hbar \omega = 3.1$~eV  to a logarithmic plot of the electrical field magnitude obtained by the numerical calculations. In general, a very good agreement between simulation and experimental photoemission pattern is achieved.
For a horizontally aligned electrical field vector pointing along the nearest neighbor direction a hexagonal lattice of dots is visible in the photoemission pattern, highlighted by the blue boxes as a guide to the eye. The simulated near-field intensity distribution yields locally confined spot-like maxima exhibiting the same hexagonal symmetry as seen in the experiment inside the material located between neighboring holes connected by the direction of the E-field vector. 
A counter-clockwise rotation of the polarization axis by 30$^\circ$ yields a completely different photoemission pattern. Instead of hexagonal dots we observe a broader distributed stripe pattern. Once again, the simulation is in good agreement showing broad linear maxima of the electric field between the holes. The stripes are aligned perpendicular to the electric field vector.
When the polarization axis is rotated by a total of 60$^\circ$, corresponding to the symmetry of the hexagonal anti-dot lattice of the sample, the photoemission image shows again a spot-like pattern. Comparing the spot-like pattern to the first image, one can clearly see a shift in position (see the blue boxes inserted into the images as a guide to the eye). This is in accordance with the simulated near-field distribution which shows the localized spots between neighboring holes along the direction of the E-field vector. The shift in position is directly reproduced.
Due to the symmetry of the sample, a rotation of the polarization axis by a total of 90$^\circ$ shows a similar situation as a rotation by 30$^\circ$. The photoemission pattern as well as the simulation show a stripe-like pattern, where the stripes are aligned perpendicular to the E-field vector.

\noindent In conclusion, comparing the photoemission pattern to the simulations provides the understanding of the polarization-dependent near-field distribution at plasmonic resonances. Whenever the electrical field vector is aligned parallel to the nearest neighbors ($\Gamma$-K direction) the near-field distribution shows a locally confined spot-like pattern. In contrast, when the electrical field vector is aligned along the direction of the next-nearest neighbors ($\Gamma$-M direction), the near-field is more distributed, forming lines perpendicular to the E-field vector. In the simulations, the E-field was also recorded at several heights: 5~nm above the whole structure, exactly in the middle of the 2~nm Au layer (thus, 1~nm below the sample surface) as shown in Fig.~\ref{fig:simulation} and in the middle of the 100~nm thick Ni film. There is almost no qualitative difference between the Au and Ni monitor, we see the same dot-like and stripe-like pattern there, depending on the light polarization.

\subsection{Correlation of photoemission electron microscopy and magneto-optic enhancement}

\begin{figure}[]
 \begin{center}
 \subfigure[][]{\includegraphics[width=0.48\textwidth]{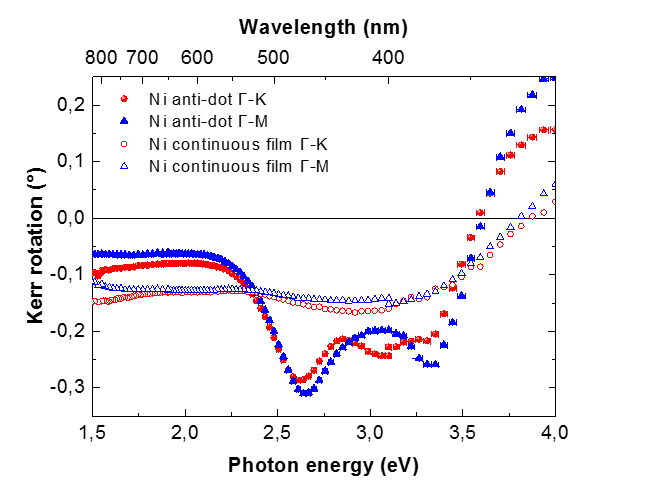}\label{fig:mokea}}
 \subfigure[][]{\includegraphics[width=0.48\textwidth]{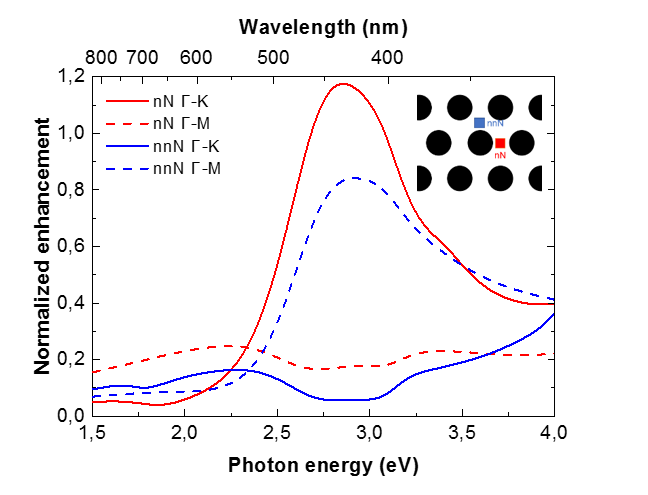}\label{fig:mokeb}}
  \end{center}
  \caption{(a) Polar Kerr rotation spectra of the Ni anti-dot (filled symbols) and a Ni continuous film (open symbols) used as a reference. Measurements are performed for $0^\circ$- and $90^\circ$-polarized light aligned along the $\Gamma$-K and $\Gamma$-M direction, respectively, for both samples at magnetic saturation at $\mathrm{B}=1$~T. (b) Enhancement of the electric near-field intensities normalized to the incident field value and calculated for $0^\circ$- and $90^\circ$-polarized light from the simulated patterns in Fig.~\ref{fig:simulation}. The graphical inset shows the points on the surface of the hexagonal lattice where the intensity is calculated. nN denotes the nearest neighbor position where spot-like emission has been observed for $0^\circ$ polarization ($\Gamma$-K) while nnN marks the next-nearest neighbor position where line-formation has been observed for $90^\circ$ polarization ($\Gamma$-M).}
\label{fig:moke}
\end{figure}

\noindent The dramatic changes in the near-field distribution recorded by PEEM can be correlated to the magneto-optic Kerr response of the structure. Figure~\ref{fig:moke} shows the polar Kerr rotation spectra for different polarization states of the incident light (left axis). Spectra are recorded for the anti-dot sample and simultaneously for a continuous reference film of the same thickness and composition. The Polar Kerr experiment was performed at $\theta= 4^\circ$ angle of incidence equivalent to the PEEM experiments, with the plane of incidence at $\phi=0$  and at magnetic saturation of the samples.
The magneto-optical Kerr effect appears when a p-polarized  component (or s-) is present in the reflection from incident s-polarized light (or p-). The complex Kerr rotation $\tilde{\Phi}$  can be related to the corresponding Fresnel reflection coefficients ($r_{\textrm{pp}}$, $r_{\textrm{ss}}$, $r_{\textrm{ps}}$,  $r_{\textrm{sp}}$) and for example for incident s-polarized light takes the form of~\cite{Bader94}:

\begin{equation}
\label{eq:polar}
\tilde{\Phi_{\mathrm{s}}} = \theta_{\mathrm{Ks}}  +\mathrm{i}   \eta_{\mathrm{Ks}}=\frac{r_{\textrm{ps}}}{r_{\textrm{ss}}}
\end{equation}
where  $\bf \mathrm{\eta_{\mathrm{K}}}$  is the polar Kerr ellipticity and $\bf \mathrm{\theta_{K}}$  defines the polar Kerr rotation as the tilt angle of the  polarization plane with respect to the incident light.  The value of the Kerr rotation can be  influenced by two factors\cite{Antonio,Papaioannou15, Fumagalli1996}: the intrinsic magneto-optically active electronic transitions ($r_{\textrm{ps}}$ - polarization conversion or pure magneto-optic contribution) and the extrinsic reflectivity ($r_{\textrm{ss}}$ - pure optical contribution).

\noindent Figure \ref{fig:moke} (a, open symbols) shows that the reference film exhibits two broad features around $ \hbar \omega = 1.5 $~eV and around $ \hbar \omega = 3 $~eV for both polarization states. This behavior has been associated in Ni with the d-p transitions of electrons at the top majority spin band and at the bottom of the minority spin band \cite{buschow83}. These characteristics are changed in the case of the patterned film (a, filled symbols). As Fig.~\ref{fig:moke}(a) reveals, a large enhancement of the polar Kerr rotation $\bf \mathrm{\theta_{\mathrm{K}}}$ is observed with respect to the continuous film for incident light polarized along both, $\Gamma$-K and $\Gamma$-M, directions at photon energies above $ \hbar \omega = 2.4$~eV. 
Interestingly at $ \hbar \omega = 2.4$~eV, the values of the Kerr rotation coincide for both polarization directions and for the reference continuous sample. At this energy, the theoretical calculations predict no surface plasmon resonance and the PEEM images show no polarization-dependent features. Above this photon energy the values of the Kerr rotation are enhanced and remain higher compared to the continuous film up to a photon energy of $ \hbar \omega = 3.5$~eV. At even higher photon energies the Kerr rotation changes sign but still remains higher in absolute values than that of the continuous film for both polarizations. 
 
\noindent PEEM images recorded at $ \hbar \omega = 3.1$~eV and $ \hbar \omega = 3.4$~eV have revealed the distribution of the near-field either as stripe-lines or as bright spots depending on the polarization state. This spatial distribution of the electric near-field recorded by PEEM influences the magneto-optic enhancement. We have determined the intensity of the enhanced electric field patterns with the help of simulations in Fig.~\ref{fig:simulation} for the full spectral range. The corresponding curves are plotted  in Fig.~\ref{fig:moke}(b) for a polarization orientation along $\Gamma$-K and $\Gamma$-M, each calculated in a position where an enhanced photoelectron yield has been observed for one of those orientations: between two neighboring holes (denoted as nN, see the inset graphic), where spot-like emission has been observed, and between next-nearest neighbors (denoted as nnN), where the stripe-formation has been observed.
We see that the electric fields are strongly intensified for light polarized either along the $\Gamma$-K or $\Gamma$-M direction (the first possesses higher intensity) at a spectral region between 2.5 and 3.5 eV. Inside this spectral region we also observed the strong magneto-optic enhancement.  This is a clear indication that the intensification of the electric field generated by the spatial distribution of the surface plasmons leads to an increase of the pure magneto-optic contribution. In particular $r_{\textrm{sp}}$ can be written as \cite{Antonio}:

\begin{equation}
\label{eq:fresnel}
\lvert {r_{\textrm{sp}}} \rvert \propto \langle E_{\textrm{p}} E_{\textrm{s}} \rangle d \lvert\epsilon_{\textrm{mo}}\rvert
\end{equation}

\noindent where $d$ is the thickness of the film, $\epsilon_{\textrm{mo}}$  its magneto-optical constant and  $\langle E_{\textrm{p}} E_{\textrm{s}} \rangle$ is the mean value of the product of both components of the field inside the MO layer. The enhancement of both components of the electric field shown in Fig.~\ref{fig:mokeb} increases the pure magneto-optic contribution of the Kerr rotation. The enhancement of the electric field is also present for energies above $ \hbar \omega = 3.5$~eV leading also to higher positive values of the Kerr rotation.

\begin{figure}[]
 \begin{center}
 \includegraphics[width=10 cm]{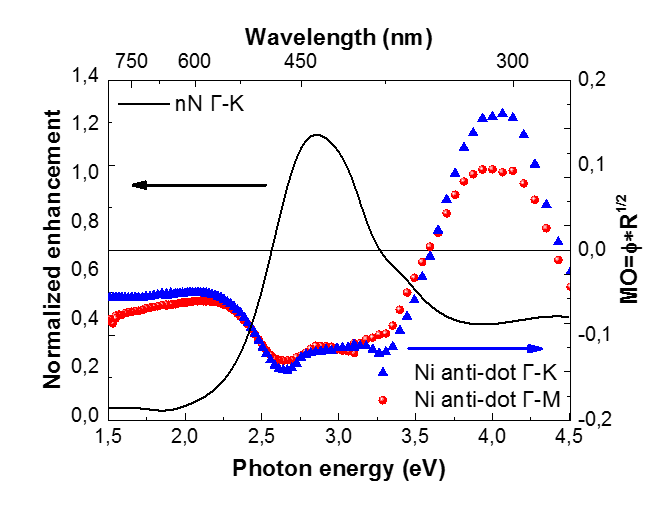}
  \end{center}
  \caption{(Left axis, black line) Enhancement of the electric field simulated for polarization along $\Gamma$-K direction, see Fig. \ref{fig:mokeb}. (Right axis, symbols) Magneto-optic figure of merit calculated for both polarization states. The reflectivity is responsible for the spectral differences between incident polarization along $\Gamma$-K and $\Gamma$-M direction. However, the Kerr enhancement is defined by the magnitude of the electric field.}
\label{fig:FOM}
\end{figure}

\noindent Still, the good agreement between  intensification of electric field and large enhanement for both polarizations of incident light can not fully explain  all the observed spectral differences between light polarized along the $\Gamma$-K and $\Gamma$-M direction.  
Characteristic features between $ \hbar \omega = 2.4-3.5 $~eV photon energies are observed: the  use of light polarized along the $\Gamma$-M direction exhibits slightly higher Kerr rotation up to  $ \hbar \omega = 2.9$~eV compared to the $\Gamma$-K direction. Around $ \hbar \omega = 3.1$~eV the signal decreases and light polarized along the $\Gamma$-K direction exhibits higher Kerr signal. Close to the plasmonic resonance of $ \hbar \omega = 3.4$~eV the light polarized, again, along the $\Gamma$-M direction exhibits bigger values than along the $\Gamma$-K direction. These differences are due to the variation of reflectivity as shown in Fig.~\ref{fig:reflectivity}. As discussed above, the Kerr rotation also depends on the pure optical contribution. Due to the slightly different angle of incidence the precise comparison of reflectivity and Kerr spectra is difficult,  however approximately still valid. As shown in Fig.~\ref{fig:reflectivity}, the reflectivity which enters as a denominator in the expression of the Kerr rotation in Eq.~\ref{eq:polar} displays a local maximum around 3.1 eV for light polarized along the $\Gamma$-M direction while the excitation along the $\Gamma$-K direction reveals a lower value. The lower reflectivity for the excitation along the $\Gamma$-K direction results in higher Kerr rotation at this energy. To reveal this dependence we have calculated the figure of merit defined as the product of Kerr rotation $\bf \mathrm{\theta_{K}}$ and the square root of reflectivity $\sqrt{\textrm{R}}$ for both polarization states, extracting the pure magneto-optical contribution. The resulting curves shown in Fig.~\ref{fig:FOM} are falling on top of each revealing  indeed that the reflectivity is responsible for the differences between both directions. However, the overall enhancement of the Kerr rotation remains and is directly associated to the magnitude of the electric field as shown in Fig.~\ref{fig:FOM} (right axis).
  
\subsection{Conclusions}

By recording with PEEM the spatial distribution of electromagnetic fields generated by the surface plasmons and by simulating the field intensity at the location of the metal/air interface we were able to correlate the enhancement of the polar Kerr effect in a Ni based magneto-plasmonic structure. PEEM images have revealed that the use of light at specific photon energies leads to polarization-dependent distributions of the near-field in form of spots or lines. A hexagonal lattice of dots is visible in the photoemission pattern for incident light polarized along the $\Gamma$-K direction, while a broader distributed pattern of stripes aligned perpendicular to the incident electric field vector is observed for the polarization aligned along the $\Gamma$-M direction. These effects are observed only at photon energies where surface plasmons are excited. The distribution of the electric near-field results in a significant increase of the field intensity in this spectral region from 2.5 to 3.5 eV. This leads to an enhancement of the Kerr rotation due to the increase of the magneto-optical conversion. The changes of reflectivity have also to be taken into account concerning the spectroscopic differences of the Kerr rotation between incident light polarized along $\Gamma$-K and $\Gamma$-M. Manipulating the Kerr enhancement around plasmonic resonances with light polarized along different angles with respect to the lattice opens new routes for tailoring the functionality of patterned magneto-plasmonic structures.

\begin{acknowledgement}

E.Th.P. acknowledges Carl Zeiss Foundation for financial support. A.G.-M. acknowledges the Spanish Ministry of Economy and Competitiveness  (Contract No. MAT2014-58860-P), and the Comunidad de Madrid (Contract No. S2013/MIT-2740). P.T. acknowledges financial support through the Excellence Initiative (DFG/GSC 266). V.K. acknowledges financial support from the Knut and Alice Wallenberg Foundation.  We thank Piotr Patoka for the preparation of the polystyrene bead template. We gratefully acknowledge the Deutsche Forschungsgemeinschaft program SFB/TRR 173: SPIN+X.

\end{acknowledgement}


\bibliography{peem-antidot}

\end{document}